\documentclass[aps,prl,twocolumn,showpacs,superscriptaddress,groupedaddress,nofootinbib,preprintnumbers]{revtex4}  % for review and submission
\usepackage{dcolumn}   % needed for some tables
\usepackage{bm}        % for math
\usepackage{amssymb}   % for math
\usepackage{latexsym}
\usepackage{amssymb}
\usepackage{amsmath}
\usepackage{fancyhdr}
\usepackage{subfigure}
\usepackage[dvips]{graphicx}

\usepackage{amsfonts}
\usepackage{epsfig}

\usepackage{color}
\outer\def\beginsection#1\par{\medbreak\bigskip
      \message{#1}\leftline{\bf#1}\nobreak\medskip
\vskip-\parskip
      \noindent}

\newcommand{\eq}{\begin{equation}}

\newcommand{\eqx}{\end{equation}}
\newcommand{\eqn}{\begin{eqnarray}}

\newcommand{\bi}{\begin{itemize}}

\newcommand{\eqnx}{\end{eqnarray}}
\newcommand{\ei}{\end{itemize}}

\begin{document}

\preprint{DESY 2012-185}

\title{On non-trivial spectra of trivial gauge theories}
\author{Piotr Korcyl}
\affiliation{NIC, DESY Zeuthen, Platanenallee 6, D-15738 Zeuthen, Germany}
\affiliation{M. Smoluchowski Institute of Physics, Jagiellonian University, ul. Reymonta 4, 30-059 Krakow, Poland}
\author{Mateusz Kore\'{n}}
\affiliation{M. Smoluchowski Institute of Physics, Jagiellonian University, ul. Reymonta 4, 30-059 Krakow, Poland}
\author{Jacek Wosiek}
\affiliation{M. Smoluchowski Institute of Physics, Jagiellonian University, ul. Reymonta 4, 30-059 Krakow, Poland}

\pacs{03.65.-w, 11.15.Ha}

\begin{abstract}
In this Letter we point out that the analytic solution of the two dimensional $U(1)$ gauge theory, on a finite lattice,
reveals in the continuum limit the renowned Manton's spectrum of topological electric fluxes together with their effective
hamiltonian and wave functions. We extend this result for the system with strings and external charges providing also a novel
interpretation of the $\Theta$ parameter. Some further generalizations are also outlined.
%In this Letter we discuss the analytic solution of two-dimensional abelian quantum field theory on a
%periodic lattice.
%Due to periodicity the system has nontrivial dynamics. We recover the renowned Manton's
%spectrum and wavefunctions by analytically performing the continuum limit.
%We provide a novel interpretation of the Manton's theta parameter and outline some possible interesting generalizations.
%%
%%In this Letter we discuss two-dimensional, abelian quantum field theory on a spatial circle from a lattice gauge theory point of view.
%%By performing analytically the continuum limit we recover known results concerning the spectrum and wavefunctions of this system.
%%We provide a novel interpretation of the Manton's $\Theta$ parameter and outline some possible and interesting generalizations.
\end{abstract}

\maketitle

\section{Introduction}
It is usually said that two-dimensional (1+1) gauge theories are trivial. In the continuum there are no transverse degrees of freedom
to sustain any dynamics. Likewise, in the lattice formulation, the system factorizes (after gauge fixing) and a partition function reduces
to  a simple one-plaquette integral \cite{Kogut,Rothe,GrossWitten}.

This picture neglects boundary conditions. It is well known  \cite{Manton}  that, e.g.,  Quantum Maxwell
Dynamics $QMD_2$ on a circle with a circumference $L$ is not entirely trivial. There remains one degree of freedom which cannot be gauged away.
In the Coulomb gauge this is the famous, constant in space, mode $A_x(t)\equiv A(t)$. Its dynamics is given by the simple, one degree of
freedom, hamiltonian
%\eqn
%H= - \frac{e^2 L}{2} \frac{d^2}{d \chi^2}, \qquad \textrm{ with } \chi=L A.  \label{H}
%\eqnx
\eqn
H= - \frac{e^2}{2L} \frac{d^2}{d A^2}.  \label{H}
\eqnx
Remaining gauge freedom allows to bring $A$ to the interval $[ 0 , 2\pi/L )$, and to identify points $A=0$ and $A=2\pi/L$, hence
the field space is also a circle with circumference $L_A=2\pi/L$ \footnote{Our $L$ differs by a factor of $2\pi$ from that of
\cite{Manton}.}. The spectrum and periodic wave functions of the above system are simply
\begin{align}
E_n&=\frac{1}{2} e^2 n^2 L, \quad n=0,\pm 1, \pm 2,..., \label{En} \\
\psi_n(A)&=\frac{1}{\sqrt{L}} e^{i n L A} \nonumber
\end{align}
%\begin{align}
%E_n&=\frac{1}{2} e^2 n^2 L, \quad n=0,\pm 1, \pm 2,..., \label{En} \\
%\psi_n(\chi)&=\frac{1}{\sqrt{L}} e^{i n \chi} \nonumber
%\end{align}
and describe the quantized states of electric flux which wraps around the circle. These are the straightforward quantum strings, with energies
proportional to their length and the string tension $\sigma_n=\frac{e^2 n^2}{2}$. These solutions exist without any external charges, the
Gauss's law being satisfied due to the non-trivial topology of a circle. Therefore they are again the simplest examples of topological strings.

In this note we shall show that this spectrum, and the hamiltonian can be obtained also from the continuum limit of the
standard lattice formulation of $QMD_2$.
Moreover, we will derive the generalization of eq.~(\ref{H}) describing the system of topological strings together with
external charges. This will also provide a straightforward, not surprising, but hopefully novel interpretation of the Manton's
parameter $\Theta$ (see also \cite{Coleman}).

Generalizations for arbitrary number of charges as well as for the non-abelian case will be also outlined.

\section{$QMD_2$ on a lattice}
Consider an $ N_t \times N_x $ lattice with unitary variables: $ U_{l}=e^{i \theta_{l}} $ associated
with spatial link $l$ and $ U_{l}=e^{i \vartheta_{l}} $ associated
with temporal link $l$. The partition function of this, pure gauge, theory reads
\begin{align}
Z&=\int d (\textrm{links}) \prod_{\textrm{plaquettes}} B(\textrm{plaquette}), \nonumber \\
B(p)&=e^{\beta \cos{\phi_p}},
\end{align}
with plaquette angles $\phi_p = \theta_i + \vartheta_j - \theta_k - \vartheta_l$, where
$\{i,j,k,l\}$ are appropriate indices of links belonging to plaquette $p$. This integral is known exactly. Using the character expansion
for $U(1)$, changing variables to plaquette angles, gives\footnote{On two-dimensional, periodic lattices there is one linear constraint
between all $N_t N_x$ plaquettes which has to be taken into account.}
\eqn
Z=\sum_{n=-\infty}^{\infty} I_n(\beta)^{N_t N_x}. \label{Zlat}
\eqnx
To recover the continuum limit, one tunes lattice coupling  $\beta$ for each value of lattice constant $a$ according to
\eqn
\beta=\frac{1}{e^2 a^2}
\eqnx
and expresses all lattice distances in physical units, i.e. $T=a N_t$, $L=a N_x$. The natural physical unit which emerges is the
dimensionful charge $e$.

Using asymptotic form of modified Bessel functions gives, up to a constant factor,
\eqn
Z \sim \sum_n \left(e^{-\frac{n^2}{2\beta}}\right)^{N_x N_t}=\sum_n e^{-\frac{e^2 n^2 L}{2} T}
\eqnx
which proves that indeed the partition function of $QMD_2$ is saturated by topological fluxes eq.~(\ref{En}).

\subsection{Lattice transfer matrix and Feynman kernel }

It is even more  instructive to derive the above equivalence from the transfer matrix formulation.
To this end we employ the Coulomb gauge on the lattice. In fact the Coulomb condition on the
2-dimensional lattice can be satisfied exactly. Consider a row of horizontal (space-like) link angles
${\theta_1, \theta_2,...,\theta_{N_x}}$. By local gauge rotation $\alpha_i$ they transform to
\eqn
^g \theta_i \rightarrow \theta_i + \beta_i , \; \; \; \; \beta_i = \alpha_i - \alpha_{i+1},\;\;\;\Sigma_i \beta_i = 0.
\eqnx
Choosing
 \eqn
 \beta_i=\frac{1}{N_x}\sum_{j=1}^{N_x} \theta_j - \theta_i,  \label{Cg}
 \eqnx
 brings all angles to the same value $\theta=\Sigma_j \theta_j /N_x$, thereby satisfying the Coulomb condition.
This procedure fixes all links in one row to the same value, however that value is not fixed since the conditions
eq.~(\ref{Cg}) leave one rotation angle $\alpha_i$ free.

The transfer matrix in this gauge, in the angular representation, is
given by the $N_x$-fold integral over vertical (time-like) links
\eqn \langle \theta | { \cal T} | \theta' \rangle = \int \prod_{j=1}^{N_x}
d\vartheta_j \prod_p B(p),  \label{TZ} \eqnx of the product of $N_x$
Boltzmann factors corresponding to all $N_x$ plaquettes between two
nearest neighbor  rows of horizontal links. Due to our gauge choice
all plaquettes depend on the same angles $\theta$ and $\theta'$,
similarly the states depend only on one angle. Notice also, that
since we are not using the temporal gauge, the integrations over the
vertical links have to be explicitly included.

Now we use again the character expansion for Boltzmann factor and integrate over vertical links to obtain
\eqn
\langle \theta | { \cal T} | \theta' \rangle = \sum_n   I_n(\beta)^{N_x} e^{ i n N_x (\theta - \theta')}
\eqnx
which in the continuum limit becomes
\eqn
\langle \theta | {\cal T} | \theta' \rangle = \sum_n  e^{-\frac{e^2}{2} n^2 L a} e^{ i n L (A - A') } = K( A, A', \epsilon).
\eqnx
This is nothing but a propagator of a 1 DOF quantum mechanical system eq.~(\ref{H},\ref{En}) over a time $\epsilon = a$.

\section{Topological fluxes with external charges}%====================================
\subsection{Eigenenergies}%=======================================
To place on a circle two external, static charges separated by a distance $R $ consider the correlation function of two Polyakov loops,
$ \langle P(0)^{\dagger} P(n_x) \rangle $, separated by $n_x$ lattice units. Standard, lattice textbook calculation gives then
\eqn
Z \langle P(0)^{\dagger} P(n_x) \rangle = \sum_n I_n(\beta) ^{N_t(N_x-n_x)} I_{n+1}(\beta)^{N_t n_x},       \label{PPNUM}
\eqnx
which in the continuum limit reads $R=a n_x$
\eqn
Z \langle P(0)^{\dagger} P(n_x) \rangle =\sum_n e^{- E^{PP}_n T},\;\;\;\;
\eqnx
with
\begin{multline}
E_n^{PP} = \frac{e^2}{2}\left( n^2 (L- R) + (n+1)^2 R \right) =\\= \frac{e^2 (n+\rho)^2 }{2} L + \frac{e^2}{2} L \rho (1- \rho)  \label{EPP}
\end{multline}
with
\begin{equation}
\rho=\frac{R}{L} \nonumber
\end{equation}
This result has a simple and appealing interpretation. Time-like Polyakov loops modify Gauss's law at spatial points $x=0$,
and $x=R$ -- they introduce unit charges at these positions. Such charges cause additional unit of flux extending over a distance R.
Hence the two contributions to the eigenenergies: an "old" flux over a distance $L-R$ and the "new" one, bigger by one unit (fluxes are
additive), over a distance $R$.

Interesting special cases:
\begin{itemize}
\item[1.] at large $T$ lowest state $n=0$ dominates and we have just a standard string of length $R$,
\item[2.] with $R=0$ the $n$-th energy level is given by topological flux of charge $n$,
\item[3.] with $R=L$, i.e. when two charges meet at the "end points" of a circle,
they annihilate ($e^+\delta_{periodic}(x) + e^-\delta_{periodic}(x-L))=0$)
and leave behind a topological string with length $L$ and charge bigger by one unit. In other words the energy levels are shifted $n \rightarrow n+1$.
\end{itemize}
 Hence varying the distance $R$ allows us to interpolate between integer valued topological fluxes. This is the meaning of the second
 representation in eq.~(\ref{EPP}). However the first term is not the whole story (apart from $\rho=0,1$). There is also the second, constant
 in $n$ but $R$ and $L$ dependent, contribution which guarantees the linear dependence of eigenenergies on distances involved.
\subsection{The hamiltonian}%====================================
The transfer matrix corresponding to eq.~(\ref{PPNUM}) is similar to eq.~(\ref{TZ})
\begin{equation}
\langle \theta | { \cal T}_{PP} | \theta' \rangle = \int \prod_{j=1}^{N_x} d\vartheta_j \prod_p B(p) e^{-i \vartheta_1} e^{ i \vartheta_{n_x+1}}
\end{equation}
except for two additional link variables coming from Polyakov lines $P(1)$ and $P(n_x+1)$. Again we have chosen the Coulomb gauge, hence
the matrix element depends only on two angles $\theta$ ($\theta'$) which specify a state of upper (lower) row. As before, upon expanding
in characters and integrating over vertical links, one obtains
\begin{multline}
\langle \theta | { \cal T}_{PP} | \theta' \rangle =\\= \sum_n  I_n(\beta)^{N_x-n_x} I_{n+1}(\beta)^{n_x}  e^{i n N_x (\theta - \theta')}
e^{i n_x (\theta - \theta')}
\end{multline}
which in the continuum limit reads
\begin{multline}
\langle \theta | { \cal T}_{PP} | \theta' \rangle = \sum_n  e^{-\frac{e^2 L a}{2} \left((n+i\rho)^2+\rho(1-\rho)\right)} e^{ i (n+\rho) L (A - A') }
\\= \tilde{K}_{PP}(A,A',a),\;\;\;\rho=\frac{R}{L}.
\end{multline}
This is again a simple Feynman kernel propagating a one DOF system by a time lapse $a$.

An explicit form of  the corresponding hamiltonian depends on the basis of eigenfunctions we choose.
One possibility is
\begin{equation}
\tilde{H}_{PP} =  \frac{e^2 L }{2} \Big( -\frac{d^2}{d \chi^2} + \rho (1 - \rho)\Big),
\end{equation}
\begin{equation}
\psi_n(\chi) = e^ {i (n+\rho) \chi},\;\;\;\chi=L A \nonumber
\end{equation}
in this case wave functions are not periodic. The other choice is
\begin{equation}
H_{PP} = \frac{e^2 L }{2} \Big( -\Big(\frac{d}{d \chi} + i \rho \Big)^2 + \rho (1 - \rho) \Big),
\end{equation}
\begin{equation}
\psi_n(\chi) = e^ {i  n  \chi} \nonumber
\end{equation}
with periodic eigenfunctions. The two are related by the transformation $\psi \rightarrow e^{i\rho \chi}\psi$, which in general is not periodic.

This freedom corresponds exactly to the ambiguity discussed in
\cite{Manton} and \cite{Asorey}, with Manton's parameter $\Theta$ acquiring now a
straightforward interpretation \eqn \Theta=\frac{R}{L}.
\label{theta} \eqnx
Namely,  as always said, $\frac{e^2}{2}\Theta$ represents the external electric field in this context.
In the original theory, in the finite volume $L$, the field extends over a fraction  $R/L$ of a whole volume.
However in our  1 DOF system the notion of the spatial distance is lost. Therefore (\ref{theta}) represents an effective field in
the reduced model, i.e. the field in the extended $QMD_2$, but {\em averaged} over the whole volume $L$.
%
%Namely,  as always said, $\frac{e^2}{2}\Theta$
%represents the external electric field in this context. In the
%original theory, in the extended volume $L$, this is obviously the
%field generated by two static charges separated by $R$. However in
%our  1 DOF system the notion of the spatial distance is lost.
%Therefore eq.~(\ref{theta}) represents an effective field in the reduced
%model, i.e. the  original field in the extended $QMD_2$, but {\em
%per unit} of the original, extended volume $L$.
%
In fact the equivalence discussed above is the simplest example of the
dimensional reduction so successful in many studies \cite{Bjorken,
BrinkSchwarz,CloudsonHalpern,Luscher,BFSS,JW,Korcyl,
Trzetrzelewski}.

\section{Generalizations and summary}%=================================
Two extensions immediately suggest themselves.

One, is to add many different, static charges (with total charge being zero) in various positions. Corresponding lattice correlation
functions of many time-like Polyakov loops can be readily calculated analogously to eq.~(\ref{PPNUM}). For example four charges with
different magnitudes will be described by
\begin{multline}
Z \langle P(i)^{\dagger} P(j)^{2\dagger} P^2(j+n_2)P(i+n_1) \rangle =\\= \sum_n I_n(\beta) ^{N_t(N_x-n_1-n_2)}
I_{n+1}(\beta)^{N_t (n_1-n_2)} I_{n+3}(\beta)^{N_t n_2}, \nonumber
\end{multline}
with doubly charged sources located inside the single charged ones, i.e. $ R_2 \le R_1 \le L$.
Repeating above calculations leads to the following eigenenergies in the continuum limit
\begin{multline}
E_n^{PP} = \frac{e^2}{2} \big( n^2 (L- R_1-R_2) + \\+ (n+1)^2 (R_1-R_2)  +(n+3)^2 R_2 \big) \nonumber
\end{multline}
etc. Corresponding 1 DOF quantum mechanical systems can be also readily constructed. This time $\Theta=(R_1+2R_2)/L$, i.e. it is
again equal to the external field averaged over the whole volume.

Second generalization is for the non-abelian pure gauge theory, $QYMD_2$, with arbitrary number of colors.
The lattice solutions eq.~(\ref{Zlat}) and eq.~(\ref{PPNUM}) are basically the same with Bessel functions replaced by the coefficients of
the character expansions of Boltzmann factors for an SU(N) gauge group. Corresponding continuum energies follow from the large
$\beta$ behavior  of these coefficients.

To summarize, the spectrum of topological fluxes, predicted by Manton quite some time ago, can be also obtained from the continuum
limit of seemingly trivial two-dimensional, abelian lattice gauge theory. The hamiltonian of the corresponding reduced
system also follows from the lattice formulation in Coulomb gauge. Addition of external charges on a lattice leads again to a
simple 1 DOF quantum system.  Resulting hamiltonian is, up to a new constant term, the same as Manton's one with non-zero
$\Theta$ parameter, which acquires a straightforward interpretation $\Theta=R/L$. Generalizations for many charges and for
non-abelian theories were also briefly outlined. We plan to discuss these issues in more detail elsewhere.

\begin{acknowledgements}
This work was partially supported through NCN grant nr 2011/03/D/ST2/01932, by Foundation for Polish Science MPD Programme co-financed
by the European Regional Development Fund, agreement no. MPD/2009/6 and by Faculty Grant no. DSC/000700/2012.
\end{acknowledgements}


\begin{thebibliography}{99}
\bibitem{Kogut} T. Banks, J. Kogut, L. Susskind, \emph{Strong-coupling calculations of lattice gauge theories: (1 + 1)-dimensional exercises}, Phys. Rev. D 13 (1976) 1043,
%A. Casher, J. Kogut, L. Susskind, \emph{Vacuum polarization and the absence of free quarks}, Phys. Rev. D 10 (1974) 732,
%J. Kogut, L. Susskind, \emph{The parton picture of elementary particles}, Phys. Rep. 8 (1973) 75,
\bibitem{Rothe} H. J. Rothe, \emph{Lattice gauge theories: an introduction}, World Scientific Lecture Notes in Physics vol.74 (2005),
\bibitem{GrossWitten} D. J. Gross, E. Witten, \emph{Possible third-order phase transition in the large-N lattice gauge theory}, Phys. Rev. D 21 (1980) 446,
\bibitem{Manton} N.S. Manton, \emph{The Schwinger model and its axial anomaly}, Ann. Phys. 159 (1985) 220,
\bibitem{Coleman} S. Coleman, R. Jackiw, L. Susskind, \emph{Charge Shielding and Quark Confinement in the Massive Schwinger Model}, Annals Phys. 93 (1975) 267,
\bibitem{Asorey} M. Asorey, J.G. Esteve, A.F. Pacheco, \emph{Planar rotor: The theta-vacuum structure, and some approximate methods in quantum mechanics}, Phys Rev D 27 (1983) 1852,
\bibitem{Bjorken} J.D. Bjorken, \emph{Elements of quantum chromodynamics}, SLAC-PUB-2372 (1979),
\bibitem{BrinkSchwarz} L. Brink, J. Schwarz, J. Scherk, \emph{Supersymmetric Yang-Mills theories}, Nucl. Phys. B 121 (1977) 77,
\bibitem{CloudsonHalpern} M. Claudson, M.B. Halpern, \emph{Supersymmetric ground state wave functions}, Nucl. Phys. 250 (1985) 689,
\bibitem{Luscher} M. L\"uscher, \emph{Some analytic results concerning the mass spectrum of Yang-Mills gauge thoeries on a torus}, Nucl. Phys. B 219 (1983) 233,
%M. L\"uscher, G. M\"unster, \emph{Weak-coupling expansion of the low-lying energy values in the $SU(2)$ gauge theory on a torus}, Nucl. Phys. B 232 (1984) 445,
\bibitem{BFSS} T. Banks, W. Fischler, S. Shenker, L. Susskind,\emph{M-theory as a matrix model: a conjecture}, Phys. Rev. D 55 (1997) 6189,
\bibitem{JW} J. Wosiek, \emph{Spectra of supersymmetric Yang-Mills quantum mechanics}, Nucl. Phys. B 644 (2002) 85,
%J. Wosiek, \emph{On the $SO(9)$ structure of supersymmetric Yang-Mills quantum mechanics}, Phys. Lett. B 619 (2005) 171,
%M. Campostrini, J. Wosiek, \emph{High precision study of the structure of D=4 supersymmetric Yang-Mills quantum mechanics}, Nucl.Phys. B703 (2004) 454,
%M. Campostrini, J. Wosiek, \emph{Exact Witten index in D = 2 supersymmetric Yang-Mills quantum mechanics}, Phys.Lett. B550 (2002) 121,
%R.A. Janik, J. Wosiek, \emph{Towards the lattice study of M theory}, Nucl.Phys.Proc.Suppl. 94 (2001) 711,
\bibitem{Korcyl} P. Korcyl, \emph{Gauge invariant plane-wave solutions in supersymmetric Yang-Mills quantum mechanics}, J.Math.Phys. 52 (2011) 042102,
%P. Korcyl, \emph{ Solutions of D=2 supersymmetric Yang-Mills quantum mechanics with SU(N) gauge group}, J.Math.Phys. 52 (2011) 052105,
%P. Korcyl, \emph{Analytic calculation of Witten index in D=2 supersymmetric Yang-Mills quantum mechanics}, J. Math. Phys. 53, 102102 (2012)
\bibitem{Trzetrzelewski} M. Trzetrzelewski, \emph{The study of SU(3) super Yang-Mills quantum mechanics}, PoS LAT2005 (2006) 275.
%M. Trzetrzelewski, \emph{Large N behavior of two dimensional supersymmetric Yang-Mills quantum mechanics}, J.Math.Phys. 48 (2007) 012302,
%M. Trzetrzelewski, \emph{The Number of gauge singlets in supersymmetric Yang-Mills quantum mechanics}, Phys.Rev. D76 (2007) 085012,
%M. Trzetrzelewski, \emph{The Hamiltonian study of supersymmetric Yang-Mills quantum mechanics}, Nucl.Phys.Proc.Suppl. 171 (2007) 325.
\end{thebibliography}
\end{document}